\documentclass[secnumarabic, graphics,floatfix,nofootinbib,
nobibnotes,aps,prd,longbibliography]{revtex4-1}

\global\arraycolsep=2pt
\usepackage{stmaryrd}
\usepackage{amsmath}
\usepackage{amssymb}
\usepackage{graphicx}
\usepackage{textcomp}
\usepackage{calrsfs}
\usepackage{yfonts}
\usepackage{bm}
\usepackage{xcolor}
\usepackage{braket}
\newcommand{\ben}{\begin{equation*}}
\newcommand{\een}{\end{equation*}}
\newcommand{\bean}{\begin{eqnarray*}}
\newcommand{\eean}{\end{eqnarray*}}

\newcommand{\be}{\begin{equation}}
\newcommand{\ee}{\end{equation}}
\newcommand{\bea}{\begin{eqnarray}}
\newcommand{\eea}{\end{eqnarray}}

\newcommand{\tr}{\text{tr}}
\newcommand{\TE}{\text{TE}}
\newcommand{\DD}{\text{D}}

\begin{document}
\title{Casimir forces in inhomogeneous media: renormalization and the principle of virtual work}

\author{Yang Li}
  \email{liyang@ou.edu}
  \affiliation{H. L. Dodge Department of Physics and Astronomy, University of Oklahoma, Norman, OK 73019 USA}
\author{Kimball A. Milton}
  \email{kmilton@ou.edu}
  \affiliation{H. L. Dodge Department of Physics and Astronomy, University of Oklahoma, Norman, OK 73019 USA}
\author{Xin Guo}
  \email{guoxinmike@ou.edu}
  \affiliation{H. L. Dodge Department of Physics and Astronomy, University of Oklahoma, Norman, OK 73019 USA}
\author{Gerard Kennedy}
  \email{g.kennedy@soton.ac.uk}
  \affiliation{School of Mathematical Sciences, University of Southampton, Southampton, SO17 1BJ, UK}
\author{Stephen A. Fulling}
  \email{fulling@math.tamu.edu}
  \affiliation{Departments of Mathematics and Physics, Texas A\&M University, College Station, TX 77843-3368, USA}

\begin{abstract}
We calculate the Casimir forces in two configurations, namely, three parallel dielectric slabs and a dielectric slab between two perfectly conducting plates, where the dielectric materials are dispersive and inhomogeneous in the direction perpendicular to the interfaces. A renormalization scheme is proposed consisting of subtracting the effect of one interface with a single inhomogeneous medium. Some examples are worked out to illustrate this scheme. Our method always gives finite results and is consistent with the principle of virtual work; it extends the Dzyaloshinskii-Lifshitz-Pitaeveskii force to inhomogeneous media.
\end{abstract}
\date\today
\maketitle

\section{Introduction}
\label{Itr}
\par Casimir demonstrated in 1948~\cite{Casimir1948} that zero-point energy could have measurable effects. The Casimir effect refers to phenomena resulting from the nontrivial vacuum state of the quantum fields in the presence of external conditions, such as boundaries, nontrivial topology, varying background potentials, and curved space. Such have been intensively investigated, both theoretically~\cite{Lifshitz1956,Dzyaloshinskii1961,Milton2001,Bordag2009,Dalvit2011} and experimentally~\cite{Derjaguin1956,Black1960,Anderson1970,Sabisky1973,Lamoreaux1997,Chen2004,Decca2005,Munday2009,Klimchitskaya2009,Sushkov2011,Garrett2018,Somers2018}. There are many potentially important applications in various areas~\cite{Ball2007,Capasso2007,Rodriguez2011,Zou2013,Tang2017}.

\par In Casimir's original configuration, two infinitely large parallel perfectly conducting plates are separated by a distance $a$ in the vacuum, which gives rise to a finite force per unit area on the plate\footnote{We use the natural units $\hbar=\varepsilon_0=\mu_0=c=1$ throughout this paper.}, namely the famous Casimir force
\begin{eqnarray}
\label{eqItr.1}
\mathcal{F}=-\frac{\pi^2}{240a^4},
\end{eqnarray}
 where the negative sign signifies its attractiveness. Lifshitz~\cite{Lifshitz1956} then generalized this model to the more physical one of two parallel homogeneous dielectric media separated by vacuum. Later, Dzyaloshinskii et al.~\cite{Dzyaloshinskii1959,Dzyaloshinskii1961} (DLP) introduced another homogeneous medium as the intervening material replacing the vacuum; their results have been demonstrated experimentally~\cite{Anderson1970,Sabisky1973}. A natural next generalization is the evaluation of Casimir forces in configurations where the media are inhomogeneous~\cite{Philbin2010,Goto2012,Xiong2013,Simpson2013,Bao2016}. However, progress in that direction has been extremely slow in the last sixty years for various reasons, of which the following two are the most significant.

\par First, it is not trivial to justify the statement that Casimir forces in inhomogeneous media are well defined. It is generally known that a force $F$ acting on a body could be expressed in terms of the energy variation $\delta E$ due to the variation $\delta a$ in the body's configuration as $F=-\delta E/\delta a$. This is known as the energy-force balance relation or the principle of virtual work (PVW). Any physically acceptable scheme to calculate a conservative force should satisfy this relation. However, as shown in Ref.~\cite{Estrada2012,Fulling2013}, an ultraviolet cutoff yields an inconsistent energy-pressure relation, which they called the ``pressure anomaly," while point-splitting regularization in a neutral direction leads to plausible results~\cite{Fulling2012}. The hope of resolving this paradox motivated the replacement of sharp boundaries by steeply rising potential barriers~\cite{Milton2011,Bouas2012,Murray2016,Milton2016,Fulling2018}, and hence to the consideration of inhomogeneous dielectric media as in our current project~\cite{Parashar2018}. After renormalization, the PVW is always satisfied in the Casimir configuration and those considered by Lifshitz and Dzyaloshinskii et al. But there is no obvious proof, or even statement, of the PVW in inhomogeneous cases. For instance, because of the inhomogeneity, it is not clear how to define the energy variation induced by the virtual displacement of the boundary between two media. Any acceptable method of calculating the Casimir force in inhomogeneous media must be consistent with the satisfaction of the PVW.

\par Second, even if the Casimir force in inhomogeneous media is well defined, there remains the problem of how to extract finite terms, whose physical meanings are unambiguous, from the energy and stress tensor. Casimir had already clearly realized that some sort of subtraction or regularization is required to obtain finite results, which are not ``divergent and devoid of physical meanings"~\cite{Casimir1948}, from the summation of the zero-point energy of all the modes, $\frac{1}{2}\sum\hbar\omega$. Since then, several approaches have been adopted to regularize the vacuum energy or stress tensor, such as the ultraviolet cutoff method~\cite{Bordag2009,Fulling2012}, zeta-function regularization~\cite{Actor1995,Elizalde1994,Bordag2009}, Laurent regularization~\cite{Goto2012}, the point-splitting method~\cite{Christensen1976,Milton2011} and dimensional continuation~\cite{Bender1994,Milton2001}. Although these techniques control the divergences, in general a divergent part must be removed. Typically, one will subtract a Green's function for the case where one homogeneous medium fills the whole space, which is sometimes named as the ``bulk contribution"~\cite{Griniasty2017,Griniasty2017-2,Parashar2018}, from the total Green's function to obtain a subtracted Green's function, a procedure occasionally called the ``Lifshitz regularization"~\cite{Xiong2013,Simpson2013}. However, when trying to calculate Casimir forces in the DLP configuration with the intervening medium being inhomogeneous, the authors of Ref.~\cite{Philbin2010,Xiong2013} ruled out the feasibility of the Lifshitz regularization and introduced another one, which resulted in divergences on the boundaries with the homogeneous media, an outcome they considered to fall ``outside the current understanding of the Casimir effect." Another attempt to regularize the inhomogeneous medium was carried out by Simpson et al.\ in Ref.~\cite{Simpson2013}, using a modified Lifshitz regularization based on a piecewise homogeneity approximation. They concluded that their piecewise method is not likely to give the correct solution. Though there are many illuminating endeavors,  more effort is still needed to find the proper renormalization methods for the inhomogeneous cases.

\par In Sec.~\ref{RA}, we demonstrate the validity of calculations for Casimir forces in the DLP configuration with the media being inhomogeneous (generalized Lifshitz configuration, GLC) and in the Casimir configuration with the intervening medium being inhomogeneous (generalized Casimir configuration, GCC). A renormalization scheme based on subtraction of the force or energy of a reference configuration is also described. This method always gives Casimir forces that are finite, as shown generally with the WKB approximation, and satisfy the PVW. Our method is consistent with the well-known homogeneous results. In Sec.~\ref{EXP}, some exactly solvable examples are provided. In Sec.~\ref{CL}, we offer concluding remarks and point out possible directions for further study. In Appendices A--F, we provide mathematical details about our theoretical calculations. In~\ref{PVWFP}, we demonstrate the PVW in flat spacetime with a plane boundary. In~\ref{appT}, we use the Green's function method to calculate the vacuum expectation values of the energy and stress tensor; explicit formulas in planar geometry are given in~\ref{appT.GA}. A full presentation of the renormalization scheme can be found in~\ref{appT.RS}. The WKB argument to show the results are finite is provided in~\ref{appT.RS.GB}. Finally, \ref{appT.RS.ASE} contains details of the exactly solvable examples discussed in Sec.~\ref{EXP}.

\section{Results and Analyses}
\label{RA}
\begin{figure}
  \centering
  \includegraphics[width=0.90\textwidth]{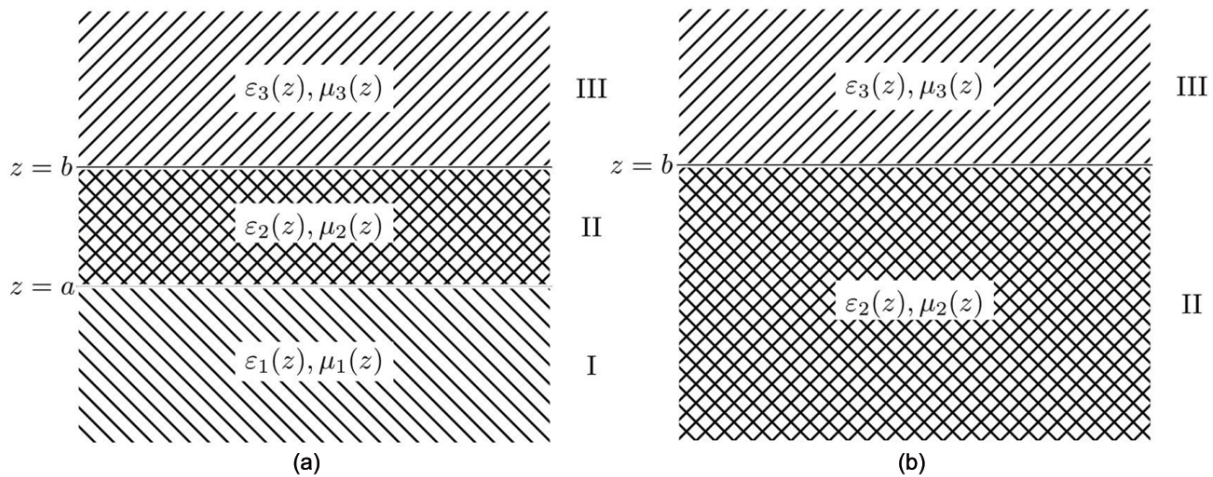}
  \caption{\label{Fig1} (a) The generalized Lifshitz configuration, where the permittivities and permeabilities of the three parallel dielectric slabs are $\varepsilon_,\mu_i,\ i=1,2,3$. (b) The reference configuration of (a) for the $z=b$ interface.}
\end{figure}
\par In this paper, we calculate the Casimir force in the configuration shown in Fig.~\ref{Fig1}a, where three parallel slabs are all isotropic, dispersive, and inhomogeneous in the $z$-direction, with the permittivity and permeability of the system $\varepsilon$ and $\mu$ being of the forms
\begin{eqnarray}
\label{eqRA.1}
\varepsilon(\zeta,z)=
\left\{
\begin{array}{ll}
  \varepsilon_3(\zeta,z), & z>b, \\
  \varepsilon_2(\zeta,z), & a<z<b, \\
  \varepsilon_1(\zeta,z), & z<a,
\end{array}
\right.
\quad
\text{and}
\quad
\mu(\zeta,z)=
\left\{
\begin{array}{ll}
  \mu_3(\zeta,z), & z>b, \\
  \mu_2(\zeta,z), & a<z<b, \\
  \mu_1(\zeta,z), & z<a.
\end{array}
\right.
\end{eqnarray}
The differential equations
\begin{eqnarray}
\label{eqRA.2}
\bigg[\partial_z\frac{1}{(\mu_i,\varepsilon_i)}\partial_z-(\varepsilon_i,\mu_i)\zeta^2-\frac{k^2}{(\mu_i,\varepsilon_i)}\bigg](\hat{e}_{i\pm},\hat{h}_{i\pm})(\zeta,\mathbf{k};z)=0,\quad i=1,2,3,
\end{eqnarray}
have solutions $\hat{e}_{i\pm}$ and $\hat{h}_{i\pm}$ satisfying proper boundary conditions, typically $\lim\limits_{z\rightarrow\pm\infty}\hat{e}_{i\pm}(z)=\lim\limits_{z\rightarrow\pm\infty}\hat{h}_{i\pm}(z)=0$.
We find, according to Appendix~\ref{appT.GA}, the transverse electric (TE) contribution to the total energy depending on the interfaces of the media is
\begin{eqnarray}
\label{eqRA.3}
\Delta U^E
&=&
\frac{1}{2}\int\frac{d\zeta d^2k}{(2\pi)^3}\ln\Delta^E(a,b),
\end{eqnarray}
with $\Delta^E(a,b)$ being
\begin{eqnarray}
\label{eqRA.3add}
\Delta^E(a,b)=[\hat{e}_{1-},\hat{e}_{2+}]_{\mu}(a)[\hat{e}_{2-},\hat{e}_{3+}]_{\mu}(b)-[\hat{e}_{1-},\hat{e}_{2-}]_{\mu}(a)[\hat{e}_{2+},\hat{e}_{3+}]_{\mu}(b),
\end{eqnarray}
where the expression $[e_i,e_j]_{\mu}(x)$ is defined as
\begin{eqnarray}
\label{eqRA.3add.20181213}
[e_i,e_j]_{\mu}(x)\equiv \frac{e'_i(x)}{\mu_i(x)}e_j(x)-e_i(x)\frac{e'_j(x)}{\mu_j(x)},
\end{eqnarray}
while the TE contribution to the discontinuity of the normal-normal stress tensor across the two sides of the interface $z=b$, i.e., $T^E_{zz}(b_\pm)$, in which $b_{\pm}=b\pm\epsilon$ and $0<\epsilon\rightarrow0$, satisfies the relation
\begin{eqnarray}
\label{eqRA.4}
T^E_{zz}(b_-)-T^E_{zz}(b_+)
=
-\frac{1}{2}\int\frac{d\zeta d^2k}{(2\pi)^3}\frac{\partial\ln\Delta^E(a,b)}{\partial b}
=
-\frac{\partial}{\partial b}\Delta U^E
.
\end{eqnarray}
The corresponding transverse magnetic (TM) contributions are obtained by making the substitution $E\rightarrow H$, $\varepsilon\leftrightarrow\mu$ and $\hat{e}\rightarrow\hat{h}$. In light of Eq.~\eqref{eqRA.4}, we see that the principle of virtual work is true in this system, which means that the Casimir forces in this kind of system are properly defined. However, these expressions are divergent.

\par In order to extract physical results, we propose a renormalization scheme based on a reference configuration for this inhomogeneous media system. Since the interaction part of the Casimir force is related to the interaction energy between the media on the upper and lower sides, when calculating the force on the $z=b$ interface (analogous arguments apply to the $z=a$ interface), we analytically extend the intervening medium II all the way down to $z\rightarrow-\infty$, that is, material II fills the whole region $z\leq b$ (shown in Fig.~\ref{Fig1}b). The reference configuration eliminates the interaction between medium I and III. This subtraction follows the same philosophy used in deriving the TGTG formula~\cite{Kenneth2006} for two bodies in homogeneous media.  For further discussion of the uniqueness and limitations of the reference subtraction method, see Ref.~\cite{Griniasty2017,Griniasty2017-2,Milton2018cp}.

\par For the reference configuration, the TE contribution to $\Delta U^E$ and $T^E_{zz}$ above are written as
\begin{eqnarray}
\label{eqRA.5}
\Delta \tilde{U}^E
=
\frac{1}{2}\int\frac{d\zeta d^2k}{(2\pi)^3}\ln\tilde{\Delta}^E(b),\quad
\tilde{T}^E_{zz}(b_-)-\tilde{T}^E_{zz}(b_+)
=
-\frac{\partial}{\partial b}\Delta \tilde{U}^E
,
\end{eqnarray}
where $\tilde{\Delta}^E(b)=[\hat{e}_{3+},\hat{e}_{2-}]_{\mu}(b)$. To obtain the renormalized energy  and normal-normal stress tensor, we subtract the reference energy and stress tensor from those of the original configuration, i.e., $\Delta U^E_r=\Delta U^E-\Delta\tilde{U}^E$ and $T^E_{r;zz}=T^E_{zz}-\tilde{T}^E_{zz}$. The force per unit area on the interface $z=b$ is thus consistent with the PVW,
\begin{eqnarray}
\label{eqRA.6}
\mathcal{F}^E
=
-\frac{\partial}{\partial b}\Delta U^E_r
=
-\frac{1}{2}\int\frac{d\zeta d^2k}{(2\pi)^3}\frac{\partial\ln\Delta^E_r(a,b)}{\partial b}
,\quad
\Delta^{E}_r(a,b)=1-\frac{[\hat{e}_{1-},\hat{e}_{2-}]_{\mu}(a)[\hat{e}_{2+},\hat{e}_{3+}]_{\mu}(b)}{[\hat{e}_{1-},\hat{e}_{2+}]_{\mu}(a)[\hat{e}_{2-},\hat{e}_{3+}]_{\mu}(b)}.
\end{eqnarray}
The TM contribution to the corresponding force is derived with the substitution $\varepsilon\leftrightarrow\mu,\ E\rightarrow H$ and $\hat{e}\rightarrow\hat{h}$. This is all discussed in more detail in Appendix~\ref{appT.RS}.

\par As a specific illustration of our renormalization method, we have considered the case where the three slabs are all homogeneous, which gives the TE contribution to the force per unit area as follows
\begin{eqnarray}
\label{eqRA.7}
\mathcal{F}^{E}
=
-\int\frac{d\zeta d^2k}{(2\pi)^3}\frac{\kappa_2}{
d^E},\quad
d^E
=
\frac{(\mu_1\kappa_2+\mu_2\kappa_1)(\mu_3\kappa_2+\mu_2\kappa_3)}{(\mu_1\kappa_2-\mu_2\kappa_1)(\mu_3\kappa_2-\mu_2\kappa_3)}e^{2\kappa_2(b-a)}-1,
\end{eqnarray}
where $\kappa_i=\sqrt{\varepsilon_i\mu_i\zeta^2+k^2}$, and its counterpart from TM modes is derived with the substitution $\mu\rightarrow\varepsilon,E\rightarrow H$. This result exactly agrees with those in Refs.~\cite{Lifshitz1956,Dzyaloshinskii1961,Milton2001}. We have also applied our method to the generalized Casimir configuration, where two parallel perfectly conducting slabs are separated by an inhomogeneous medium, and found the forces per unit area at the $z=b$ interface, when the intervening medium is homogeneous, are
\begin{eqnarray}
\label{eqRA.8}
\mathcal{F}^E
=
\mathcal{F}^{H}
=
-\frac{\pi^2}{480\sqrt{\varepsilon_2\mu_2}}\frac{1}{(b-a)^4}
,
\end{eqnarray}
which is just the result in Eq.~\eqref{eqItr.1} as long as $\varepsilon_2=\mu_2=1$. Eq.~\eqref{eqRA.8} could also be derived by taking the limit $\mu_1=\mu_3=1$ and $\varepsilon_1,\varepsilon_3\rightarrow\infty$ in Eq.~\eqref{eqRA.7}. Therefore, our method is consistent with previous results derived in the homogeneous cases.

\par To show that our renormalized results are finite, we utilized the WKB approximation to illustrate the leading behaviors of both GLC and GCC in Eq.~\eqref{eqT.RS.GB.3} and Eq.~\eqref{eqT.RS.GB.4}. As usually expected, in the high frequency region $\zeta\rightarrow\infty$, no material could respond to the electromagnetic oscillation so rapidly as to modify the field significantly, which implies the relation $\lim\limits_{\zeta\rightarrow\infty}\varepsilon(\zeta),\mu(\zeta)=1$. Consequently, the leading terms of the total energy in the GLC and GCC from TE modes in the high frequency region are
\begin{eqnarray}
\label{eqRA.9}
\Delta U^E_{r;\text{GLC(GCC)}}(|\zeta|\approx\infty)
=
\frac{1}{2}\int_{|\zeta|\approx\infty}\frac{d\zeta d^2k}{(2\pi)^3}\ln\bigg[1-\eta_{\text{GLC(GCC)}}(\zeta)e^{-2\kappa(b-a)}\bigg]
,
\end{eqnarray}
where $\kappa=\sqrt{\zeta^2+k^2}$ and the coefficients for GLC and GCC satisfy $\lim\limits_{\zeta\rightarrow\infty}\eta_{\text{GLC}}(\zeta)=0$ and $\lim\limits_{\zeta\rightarrow\infty}\eta_{\text{GCC}}(\zeta)=1$ according to Eq.~\eqref{eqT.RS.GB.3} and Eq.~\eqref{eqT.RS.GB.4}. So in high frequency region, the GLC behaves like the vacuum everywhere, which is just as expected; while for the GCC, $\Delta U^E_{r;\text{GCC}}$ is always finite, which implies a finite Casimir force. As for the integral over $k$, similar convergence can be seen from Eq.~\eqref{eqT.RS.GB.3} and Eq.~\eqref{eqT.RS.GB.4}. This demonstrates that our method yields finite results.

\par The consistency and the effectiveness of our method give us some confidence to claim that we have found a reasonable approach to evaluate the Casimir forces in the GLC and GCC, although full confirmation from solid experimental results is still required. Perhaps a differential scheme along the lines of Ref.~\cite{Bimonte2016} could be used to observe our results. The following examples demonstrate the behaviors of the Casimir forces in inhomogeneous media.

\section{Examples}
\label{EXP}
\par
\begin{figure}
  \centering
  \includegraphics[width=0.60\textwidth]{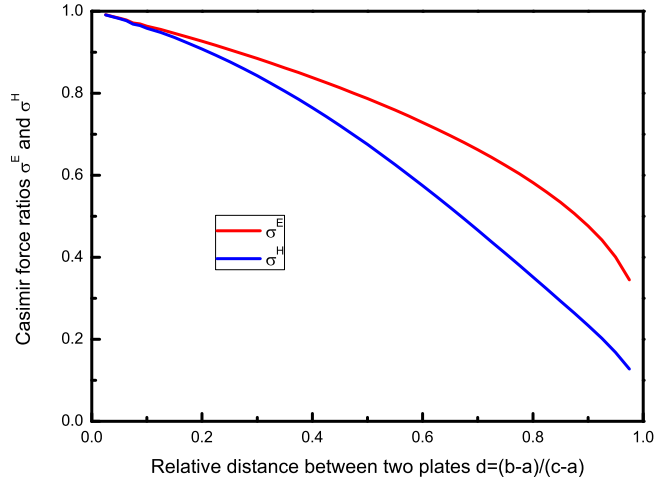}
  \caption{\label{FigPC1} The TE and TM contributions to Casimir force ratios, denoted as $\sigma^E$ and $\sigma^H$, in the GCC, where the permittivity and permeability of the intervening medium are $\varepsilon(z)=\lambda/(c-z)^2=\tilde{\lambda}/(1-d_z)^2,\ \tilde{\lambda}=\lambda/(c-a)^2,\ d_z=(z-a)/(c-a)$ and $\mu=1$ respectively, with $\tilde{\lambda}=1$. Those Casimir force ratios are defined as $\sigma^{E}=\mathcal{F}^{E}/\mathcal{F}^{\text{HE}}$ and $\sigma^{H}=\mathcal{F}^{H}/\mathcal{F}^{\text{HE}}$, where $\mathcal{F}^E$ and $\mathcal{F}^H$ are TE and TM contributions to the Casimir forces in Eq.~\eqref{eqT.RS.ASE.ISM.5} and $\mathcal{F}^{\text{HE}}$ is the homogeneous Casimir force as shown in Eq.~\eqref{eqRA.8} with permittivity and permeability being $\varepsilon(a)$ and $1$ respectively. }
\end{figure}

 There are only a few cases where the Green's functions may be explicitly constructed in terms of known functions. One of these is the
inhomogeneous medium considered in Ref.~\cite{Griniasty2017-2,Parashar2018}. First, we investigate the GCC where the permittivity and permeability of the intervening medium are $\varepsilon=\lambda/(z-c)^2$ and $\mu=1$ with $\lambda$ and $c$ as constant parameters and $b<c$. The forces are given in Eq.~\eqref{eqT.RS.ASE.ISM.5}. As a special case, for $\lambda/(c-a)^2=1$, we see in Fig.~\ref{FigPC1} how the Casimir forces from the TE and TM modes vary with the separation $d=b-a$ between two perfectly conducting plates. According to Fig.~\ref{FigPC1}, it is clear that as the distance $d$ increases, this GCC model differs significantly from the homogeneous case due to its inhomogeneity; while the GCC model converges to the homogeneous case when $d\rightarrow0$, which is intuitively reasonable since the inhomogeneity is not significant at short distances.
\par
\begin{figure}
  \centering
  \includegraphics[width=1.00\textwidth]{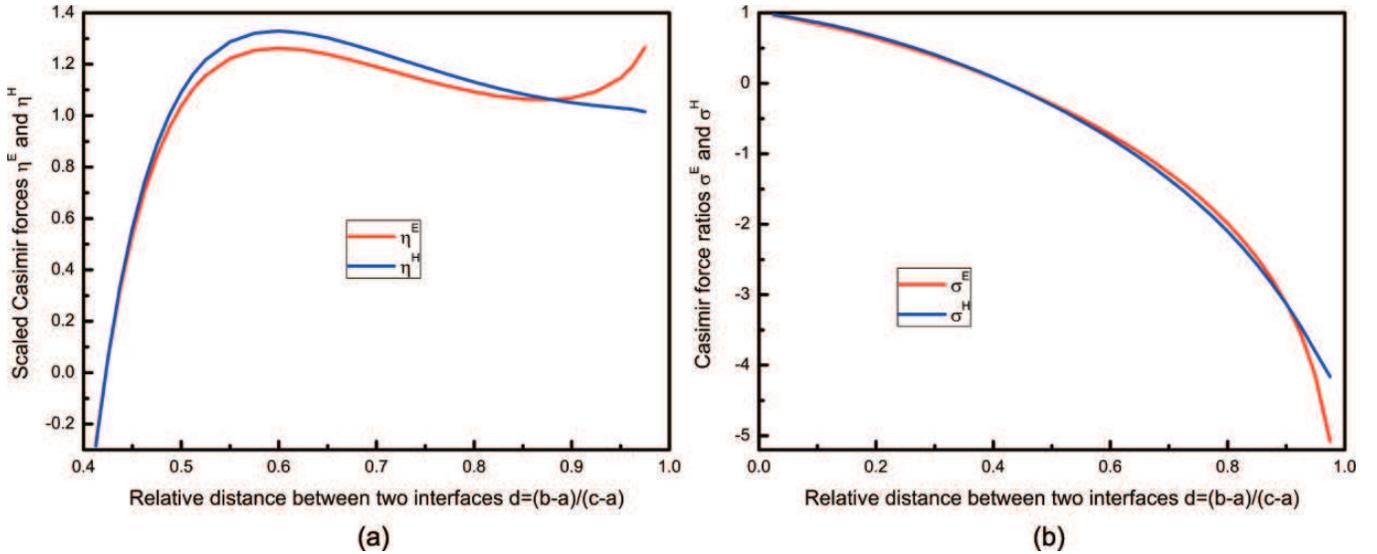}
  \caption{\label{FigDIV} The Casimir forces in the GLC, where the permittivity and permeability of the intervening medium are the same as those defined in Fig.~\ref{FigPC1}, and for the lower and upper dielectric slabs $\varepsilon_1=2,\mu_1=1$ and $\varepsilon_3=3,\mu_3=1$. (a) The TE and TM contributions to the scaled Casimir forces, defined as $\eta^E=10^{3}(c-a)^4\mathcal{F}^{E}$ and $\eta^H=10^{2}(c-a)^4\mathcal{F}^{H}$ respectively, in which $\mathcal{F}^{E}$ and $\mathcal{F}^{H}$ are given in Eq.~\eqref{eqT.RS.ASE.ISM.4}. (b) The Casimir force ratios, defined as $\sigma^{E}=\mathcal{F}^{E}/\mathcal{F}^{\text{HE}}$ and $\sigma^{H}=\mathcal{F}^{H}/\mathcal{F}^{\text{HM}}$, where the homogeneous Casimir forces $\mathcal{F}^{\text{HE}}$ and $\mathcal{F}^{\text{HM}}$ are given in Eq.~\eqref{eqRA.7} with permittivity and permeability being $\varepsilon(a)$ and $1$ respectively. }
\end{figure}

\begin{figure}
  \includegraphics[width=0.60\textwidth]{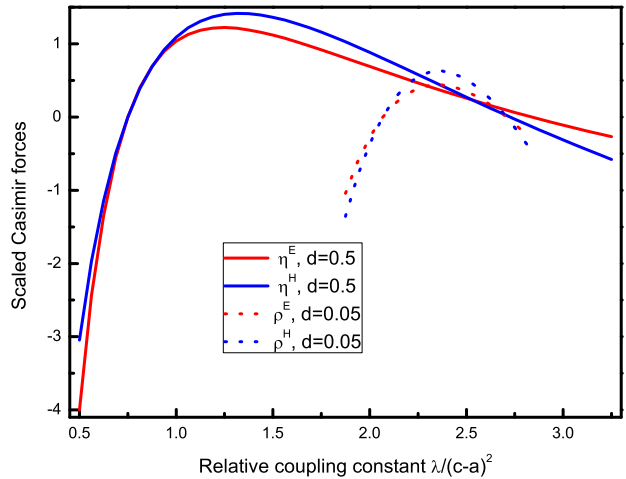}
  \caption{\label{FigCouple} Consider the same GLC configuration in Fig.~\ref{FigDIV} with the same parameters, except for $\lambda$ and $b$. The TE and TM contributions to the scaled Casimir forces, $\eta^E$ and $\eta^{H}$ as defined in Fig.~\ref{FigDIV} and $\rho^E=(c-a)^4\mathcal{F}^{E}$ and $\rho^H=10^{-1}(c-a)^4\mathcal{F}^{H}$, as functions of $\lambda$ are shown with their dependence on $\tilde{\lambda}=\lambda/(c-a)^2$. Here the $d=(b-a)/(c-a)=0.5$ case is plotted with solid lines ($\eta^E$ and $\eta^H$) and the $d=(b-a)/(c-a)=0.05$ case is plotted with dotted lines ($\rho^E$ and $\rho^H$).}
\end{figure}

 We further extend the inverse square permittivity model to the GLC case, where the dielectric slabs for the $z<a$ and $z>b$ regions are both homogeneous and the intervening medium has the permittivity and permeability as above. For the case $\varepsilon_1=2,\varepsilon_3=3$, $\mu_1=\mu_2=\mu_3=1$ and $\lambda/(c-a)^2=1$, Fig.~\ref{FigDIV}a shows the TE and TM inhomogeneous Casimir forces. Fig.~\ref{FigDIV}b shows that the separation dependence of the Casimir forces in this GLC model is distinct from that of their homogeneous counterparts in Eq.~\eqref{eqRA.7} with $\varepsilon_2=\varepsilon(a),\mu_2=1$, and that the influence of the inhomogeneity decreases as the separation between the two interfaces gets smaller. Moreover, as the interface $z=b$ is sufficiently close to $c$ the Casimir forces in this GLC will turn from attractive to repulsive. Repulsion occurs when in some region $\varepsilon_1<\braket{\varepsilon_2}<\varepsilon_3$, where $\braket{\varepsilon_2}$ is an average of $\varepsilon_2$ in some sense, as is known for the DLP configuration. Therefore, for a given separation $b-a$ and singularity position $c$, a region of $\lambda$ can be found for which the Casimir force is repulsive. For fixed $\lambda$, the TE Casimir forces do not behave monotonically in the repulsive region, see Fig.~\ref{FigDIV}a. Repulsion can occur near the $z=a$ plate when $\varepsilon_1<\lambda/(c-a)^2<\varepsilon_2$. For example see the dotted lines in Fig.~\ref{FigCouple}, where the positive force signifies repulsion of the plate at $z=b$.
\par
\begin{figure}
  \centering
  \includegraphics[width=0.50\textwidth]{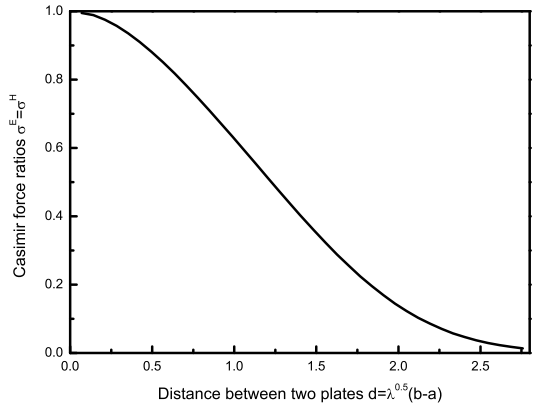}
  \caption{\label{FigDIAPH} The separation dependence of the relative Casimir force $\sigma^E=\sigma^H=\mathcal{F}^{E}/\mathcal{F}^{\text{HE}}$ in the GCC with permittivity and permeability $\varepsilon=e^{\lambda(z-c)^2}$ and $\mu=e^{-\lambda(z-c)^2}$, where $\sqrt{\lambda}(c-a)=2^{-3/2}$. $\mathcal{F}^{E}$ is given in Eq.~\eqref{eqT.RS.ASE.DM.4} and $\mathcal{F}^{\text{HE}}$ is the TE Casimir force of a homogeneous GCC satisfying $\varepsilon\mu=1$.}
\end{figure}

 To further explore the inhomogeneous effect, we calculated the Casimir forces in a GCC with a diaphanous intervening medium, meaning one whose permittivity $\varepsilon$ and permeability $\mu$ satisfy $\varepsilon\mu=1$. A diaphanous dielectric ball \cite{Brevik1982} or cylinder \cite{Milton1999} has unambiguous finite Casimir stress and energy, which without such condition would be plagued with divergences. (In the electromagnetic $\delta$-function sphere, analogous behavior was expected to be found \cite{Milton2018}, but more work is apparently needed.) Here we let the permittivity of the diaphanous medium be $\varepsilon(z)=\exp[\lambda(z-c)^2]$ and find the Casimir forces in Eq.~\eqref{eqT.RS.ASE.DM.4}. We note that $\mathcal{F}^{E}=\mathcal{F}^{M}$ is always true for this case, which is a property in common with the homogeneous cases in Eq.~\eqref{eqRA.8}. The ratio between the Casimir force in this GCC and its counterpart in the homogeneous GCC is shown in Fig.~\ref{FigDIAPH}. We see that even though the speed of light is the same as that in the vacuum, the Casimir force in this GCC is considerably different from that in the vacuum and the larger the separation the larger is the discrepancy. Of course, it reduces to the homogeneous case as the separation goes to zero. 

\section{Conclusions}
\label{CL}
\par To attain a reliable procedure for Casimir force calculations in inhomogeneous media, we have taken the first step by investigating the generalized Lifshitz configuration and the generalized Casimir configuration where the intervening media are inhomogeneous in one direction. We have proposed a renormalization scheme based on a reference configuration. This scheme is consistent with the principle of virtual work and renders the Casimir force finite for inhomogeneous and dispersive media. We have also applied our approach to a few analytically solvable examples, in which we justified the effectiveness and consistency of our method and illustrated the possibility of Casimir repulsion and nonmonotonicity in the inhomogeneous case.

\par Although our scheme always gives plausible results to date, there are still some knotty points that should be considered seriously. In particular, we have not included the interaction of one interface with the inhomogeneous medium itself. Doing so may entail understanding and modeling how realistic media behave under deformation.




\appendix
\section{Principle of Virtual Work in Flat Spacetime with a Plane Boundary}
\label{PVWFP}
\par Consider a quantized field in a static spacetime with line element $ds^2=g_{00}\,dt^2 + dx^2+dy^2+g_{zz}\,dz^2$. Under the combined coordinate scaling $t\to \alpha^{-1}\,t=t'$ and dual metric scaling $g_{00}\to \alpha^2\,g_{00}=g'_{00}$, where $\alpha>0$ is a constant scale factor, the line element, and therefore the physics, is unchanged. The corresponding invariance of the one-loop effective action, $W=\int L\,dt$, and the time independence of the one-loop effective Lagrangian, $L$, together imply that $L$ scales as $L(\alpha)=\alpha\,L(1)=\alpha L$, and therefore that $\frac{dL(\alpha)}{d\alpha}=L$. On the other hand, a small change in the scale factor results in the functional variation
$\delta L(\alpha)=\int d^3x \,\frac{\delta L}{\delta g'_{00}}\, \delta(\alpha^{2}\,g_{00})$, which implies
\begin{equation}
\label{eq:1}
\left.\frac{dL(\alpha)}{d\alpha}\right|_{\alpha=1}=\int d^3x \, \sqrt{|g|}\, g_{00}\,T^{00}=-E,
\end{equation}
where $g=\det [g_{\mu\nu}]$, and $E$ is the vacuum energy. Thus, $L=-E$. This derivation is a simplified version of that in Ref.~\cite{Dowker1978}.

\par Consider now the bounded domain $z\le b$. A virtual normal displacement of the plane boundary $z=b$ may be effected by applying the following contraction to the $z$ boundary layer $(b-h,b]$, where $h>0$ is arbitrarily small and $\beta \ge 1$ is a constant scale factor:
\begin{equation}
\label{eq:2}
\forall z \in (b-h,b], \quad z \to z'= b-h+\beta^{-1}\,(z-b+h)\in (b-h, b'],
\end{equation}
where the boundary $z=b$ maps to $z=b'\in (b-h, b]$ for $\beta=\frac{h}{b'-b+h}$.

\par On $(b-h,b]$, under the combined coordinate contraction $z \to z'$ in \eqref{eq:2} and dual metric scaling $g_{zz}\to \beta^2\,g_{zz}=g'_{zz}$, the line element, and therefore the physics, is again unchanged. The corresponding invariance of the one-loop effective Lagrangian, $L(\beta, \beta)=L(1,1)=L$, where the first argument denotes the coordinate contraction scale factor and the second argument the metric scale factor, implies that
\begin{equation}
\label{eq:3}
\left.\frac{d L(\beta,1)}{d\beta} \right|_{\beta=1^{+}}+\left.\frac{dL(1,\beta)}{d\beta}\right|_{\beta=1^{+}}=0.
\end{equation}
In effect, the combined coordinate contraction and dual metric scaling create a compound passive transformation that leaves the action form-invariant. The components of that transformation may be reinterpreted as active transformations, and the invariance may be used to relate their offsetting first-order effects. This, in essence, is the content of \eqref{eq:3}.

\par From \eqref{eq:2},
\begin{equation}
\left.\frac{d L(\beta,1)}{d\beta} \right|_{\beta=1^{+}}=-h\left.\frac{dL(\beta,1)}{db'}\right|_{b'=b^{-}}=h\,\frac{dE}{db^{-}},
\end{equation}
while,  from the functional variation
$\delta L(1,\beta)=\int d^3x \,\frac{\delta L}{\delta g'_{zz}}\, \delta(\beta^{2}\,g_{zz})$,
\begin{equation}
\left.\frac{dL(1, \beta)}{d\beta}\right|_{\beta=1^{+}}=\int_{z=b-h}^{z=b} d^3x \, \sqrt{|g|}\, g_{zz}\,T^{zz}.
\end{equation}
Thus, \eqref{eq:3} may be restated as
\begin{equation}
-\frac{d}{db^{-}}E=\frac{1}{h}\,\int_{z=b-h}^{z=b} d^3x \, \sqrt{|g|}\, g_{zz}\,T^{zz},
\end{equation}
which, in the limit $h\to 0^{+}$, becomes
\begin{equation}
-\frac{d}{db^{-}}E=\int_{z=b^-}^{} dx\,dy\, \sqrt{|g|}\, g_{zz}\,T^{zz}.
\end{equation}
Reduced to the Minkowski metric, this is a statement of the PVW for a quantized field in flat spacetime, under virtual normal displacement of a plane boundary.

\section{Formalism}
\label{appT}
\par The fundamental object in quantum field theory is the Green's function. In this appendix, we will use the Green's function to calculate the energies and stress tensors. In Euclidean spacetime, the vacuum expectation values of the dyadics of the electric and magnetic fields $\mathbf{E}$ and $\mathbf{H}$ are expressed in terms of the Green's dyadics as \cite{Milton2001}
\begin{eqnarray}
\label{eqT.1}
\braket{\mathbf{E}(x)\mathbf{E}(x')}
=
-\int\frac{d\zeta}{2\pi}e^{i\zeta(\tau-\tau')}\bm{\Gamma}_{\zeta}(\mathbf{r},\mathbf{r}'),\quad
\braket{\mathbf{H}(x)\mathbf{H}(x')}
=
-\int\frac{d\zeta}{2\pi}e^{i\zeta(\tau-\tau')}\bigg[\bm{\Phi}_{\zeta}(\mathbf{r},\mathbf{r}')
-\bm{\mu}^{-1}(\zeta,\mathbf{r})\bigg],
\end{eqnarray}
where $\tau$ is the Euclidean time, $x=(\tau,\mathbf{r})$, and the equations for the reduced Green's dyadics for each Euclidean frequency $\bm{\Gamma}_{\zeta}(\mathbf{r},\mathbf{r}')$ and $\bm{\Phi}_{\zeta}(\mathbf{r},\mathbf{r}')$ are
\begin{subequations}
\label{eqT.2}
\begin{eqnarray}
\label{eqT.2a}
\bigg[\zeta^2\bm{\varepsilon}(\zeta,\mathbf{r})+\nabla\times\bm{\mu}^{-1}(\zeta,\mathbf{r})\cdot\nabla\times\bm{1}\bigg]\cdot\bm{\Gamma}_{\zeta}(\mathbf{r},\mathbf{r}')
=\zeta^2\bm{1}\delta(\mathbf{r}-\mathbf{r}'),
\end{eqnarray}
\begin{eqnarray}
\label{eqT.2b}
\bigg[\zeta^2\bm{\mu}(\zeta,\mathbf{r})+\nabla\times\bm{\varepsilon}^{-1}(\zeta,\mathbf{r})\cdot\nabla\times\bm{1}\bigg]\cdot\bm{\Phi}_{\zeta}(\mathbf{r},\mathbf{r}')
=\zeta^2\bm{1}\delta(\mathbf{r}-\mathbf{r}'),
\end{eqnarray}
\end{subequations}
in which $\bm{\varepsilon}$ and $\bm{\mu}$ are the permittivity and permeability of the medium.

\par Suppose the medium is isotropic, dispersive, and inhomogeneous only in the $z$-direction. Then in this planar geometry the reduced Green's functions have the following forms:
\begin{eqnarray}
\label{eqT.3}
(\bm{\Gamma}_{\zeta},\bm{\Phi}_{\zeta})(\mathbf{r},\mathbf{r}')
=\int\frac{d^2k}{(2\pi)^2}e^{i\mathbf{k}\cdot(\mathbf{r}_{\parallel}-\mathbf{r}'_{\parallel})}(\mathbf{g}_{\zeta,\mathbf{k}},\mathbf{h}_{\zeta,\mathbf{k}})(z,z'),\quad
\mathbf{r}_{\parallel}=(x,y).
\end{eqnarray}
Without loss of generality, choose $\mathbf{k}$ along the $x$-axis. Then $g^E$ and $g^H$, which satisfy the equation
\begin{eqnarray}
\label{eqT.4}
\bigg[\partial_z\frac{1}{(\mu,\varepsilon)}\partial_z-(\varepsilon,\mu)\zeta^2-\frac{k^2}{(\mu,\varepsilon)}\bigg]g^{(E,H)}_{\zeta,\mathbf{k}}(z,z')=\delta(z-z'),\ k=|\mathbf{k}|,
\end{eqnarray}
are employed to express $\mathbf{g}_{\zeta,\mathbf{k}}$ as
\begin{eqnarray}
\label{eqT.5}
\mathbf{g}_{\zeta,\mathbf{k}}(z,z')
=
\left[
  \begin{array}{ccc}
     \frac{1}{\varepsilon\varepsilon'}\partial_z\partial_{z'}g^H_{\zeta,\mathbf{k}}+\frac{1}{\varepsilon}\delta(z-z') &  0 &  \frac{ik}{\varepsilon\varepsilon'}\partial_{z}g^H_{\zeta,\mathbf{k}} \\
     0 & -\zeta^2g^E_{\zeta,\mathbf{k}} & 0  \\
     -\frac{ik}{\varepsilon\varepsilon'}\partial_{z'}g^H_{\zeta,\mathbf{k}} &  0 & \frac{k^2}{\varepsilon\varepsilon'}g^H_{\zeta,\mathbf{k}}+\frac{1}{\varepsilon}\delta(z-z')  \\
  \end{array}
\right]
,
\end{eqnarray}
and $\mathbf{h}_{\zeta,\mathbf{k}}(z,z')$ is obtained with the substitution $\varepsilon\leftrightarrow\mu$ and $E\leftrightarrow H$.

\par Define functions $(e_{\pm},h_{\pm})(\zeta,\mathbf{k};z)$ as the solutions of the corresponding homogeneous differential equations
\begin{eqnarray}
\label{eqT.6}
\bigg[\partial_z\frac{1}{(\mu,\varepsilon)}\partial_z-(\varepsilon,\mu)\zeta^2-\frac{k^2}{(\mu,\varepsilon)}\bigg](e_{\pm},h_{\pm})(\zeta,\mathbf{k};z)=0
\end{eqnarray}
that satisfy the continuity conditions
\begin{eqnarray}
\label{eqT.7}
\forall z\in\mathbb{R},\ \lim_{y\rightarrow z_+}(e_{\pm},h_{\pm})(\zeta,\mathbf{k};y)=\lim_{y\rightarrow z_-}(e_{\pm},h_{\pm})(\zeta,\mathbf{k};y),\quad
\lim_{y\rightarrow z_+}\bigg(\frac{e'_{\pm}}{\mu},\frac{h'_{\pm}}{\varepsilon}\bigg)(\zeta,\mathbf{k};y)=\lim_{y\rightarrow z_-}\bigg(\frac{e'_{\pm}}{\mu},\frac{h'_{\pm}}{\varepsilon}\bigg)(\zeta,\mathbf{k};y),\quad
\end{eqnarray}
and the relevant boundary conditions, for instance $\lim\limits_{z\rightarrow\pm\infty}(e_{\pm},h_{\pm})(\zeta,\mathbf{k};z)=0$. We can then write $g^{(E,H)}_{\zeta,\mathbf{k}}(z,z')$ as
\begin{eqnarray}
\label{eqT.8}
g^{E}_{\zeta,\mathbf{k}}(z,z')=\frac{e_{+}(\zeta,\mathbf{k};z_>)e_{-}(\zeta,\mathbf{k};z_<)}{W^E_{\zeta,\mathbf{k}}},\quad
g^{H}_{\zeta,\mathbf{k}}(z,z')=\frac{h_{+}(\zeta,\mathbf{k};z_>)h_{-}(\zeta,\mathbf{k};z_<)}{W^H_{\zeta,\mathbf{k}}},
\end{eqnarray}
where the generalized Wronskians 
\begin{eqnarray}
\label{eqT.9}
W^{E}_{\zeta,\mathbf{k}}=\frac{e'_{+}e_{-}
-e_{+}e'_{-}}{\mu},\quad
W^{H}_{\zeta,\mathbf{k}}=\frac{h'_{+}h_{-}
-h_{+}h'_{-}}{\varepsilon},
\end{eqnarray}
are constant in $z$.

\par If the material has no energy and momentum dissipation, then the vacuum expectation values of the energy density $U$ and stress tensor $\mathbf{T}$ are\footnote{In the non-dissipative cases, $\varepsilon$ and $\mu$ are real.}, respectively,
\begin{subequations}
\label{eqT.10}
\begin{eqnarray}
\label{eqT.10a}
U=-\frac{1}{2}\int\frac{d\zeta}{2\pi}\bigg[\frac{d(\zeta\varepsilon)}{d\zeta}\tr\bm{\Gamma}_{\zeta}(\mathbf{r},\mathbf{r})
+
\frac{d(\zeta\mu)}{d\zeta}\tr\bm{\Phi}_{\zeta}(\mathbf{r},\mathbf{r})\bigg],
\end{eqnarray}
and
\begin{eqnarray}
\label{eqT.10b}
\bm{\mathbf{T}}
=
-\int\frac{d\zeta}{2\pi}\bigg\{
\frac{\bm{1}}{2}\tr\bigg[\varepsilon(\zeta,\mathbf{r})\bm{\Gamma}_{\zeta}(\mathbf{r},\mathbf{r})+\mu(\zeta,\mathbf{r})\bm{\Phi}_{\zeta}(\mathbf{r},\mathbf{r})\bigg]
-\varepsilon(\zeta,\mathbf{r})\bm{\Gamma}_{\zeta}(\mathbf{r},\mathbf{r})-\mu(\zeta,\mathbf{r})\bm{\Phi}_{\zeta}(\mathbf{r},\mathbf{r})\bigg\}.
\end{eqnarray}
\end{subequations}
Ignoring the unphysical divergences coming from $\delta$-functions, we may separate these into the transverse electric (TE) and transverse magnetic (TM) modes as $U=U^E+U^H,\ \mathbf{T}=\mathbf{T}^E+\mathbf{T}^H$ where
\begin{eqnarray}
\label{eqT.11}
U^{(E,H)}
=
\int\frac{d\zeta d^2k}{(2\pi)^3}u^{(E,H)}(\zeta,\mathbf{k};z),\quad
\mathbf{T}^{(E,H)}
=
\int\frac{d\zeta d^2k}{(2\pi)^3}\mathbf{t}^{(E,H)}(\zeta,\mathbf{k};z),
\end{eqnarray}
the reduced terms being
\begin{subequations}
\label{eqT.12}
\begin{eqnarray}
\label{eqT.12a}
u^E(\zeta,\mathbf{k};z)
=
\frac{-1}{2\mu W^E_{\zeta,\mathbf{k}}}\bigg[\frac{d(\zeta\mu)}{d\zeta}\frac{e'_+e'_-}{\mu}-\frac{d(\zeta\varepsilon)}{d\zeta}\mu\zeta^2e_+e_-+\frac{d(\zeta\mu)}{d\zeta}\frac{k^2}{\mu}e_+e_-\bigg],
\end{eqnarray}
and
\begin{eqnarray}
\label{eqT.12c}
\mathbf{t}^E(\zeta,\mathbf{k};z)
=
\frac{-1}{\mu W^E_{\zeta,\mathbf{k}}}
\left[
  \begin{array}{ccc}
     \frac{-e'_+e'_--\varepsilon\mu\zeta^2e_+e_-+k^2e_+e_-}{2} & 0  &  -ike'_+e_- \\
     0 & \frac{e'_+e'_-+\varepsilon\mu\zeta^2e_+e_-+k^2e_+e_-}{2}  & 0  \\
     ike_+e'_- & 0  & \frac{e'_+e'_--\varepsilon\mu\zeta^2e_+e_--k^2e_+e_-}{2}  \\
  \end{array}
\right].\quad
\end{eqnarray}
\end{subequations}
Correspondingly, $u^H,\mathbf{t}^H$ are obtained by the substitution $\varepsilon\leftrightarrow\mu,e\leftrightarrow h,E\leftrightarrow H$.

\section{Planar Geometry}
\label{appT.GA}
\par In this paper, we mainly study the system in which there are three inhomogeneous dielectric slabs at $z\leq a$, $a<z<b$ and $z\geq b$ with media whose permittivities and permeabilities, denoted $(\varepsilon_i,\mu_i),\ i=1,2,3$ respectively, are isotropic. The solutions to Eq.~\eqref{eqT.6} are given in terms of the well-defined solution in each region, i.e. $\hat{e}_{i\pm}$ and $\hat{h}_{i\pm}$, as
\begin{subequations}
\label{eqT.GA.1}
\begin{eqnarray}
\label{eqT.GA.1a}
e_{+}(z)
=
\left\{
\begin{array}{cl}
  \hat{e}_{3+}(z), & z>b, \\
  A_+\hat{e}_{2+}(z)+B_+\hat{e}_{2-}(z), & a<z<b, \\
  C_+\hat{e}_{1+}(z)+D_+\hat{e}_{1-}(z), & z<a,
\end{array}
\right.\quad
e_{-}(z)
=
\left\{
\begin{array}{cl}
  C_-\hat{e}_{3+}(z)+D_-\hat{e}_{3-}(z), & z>b, \\
  A_-\hat{e}_{2+}(z)+B_-\hat{e}_{2-}(z), & a<z<b, \\
  \hat{e}_{1-}(z), & z<a,
\end{array}
\right.
\end{eqnarray}
where the coefficients are determined by the continuity conditions,
\begin{eqnarray}
\label{eqT.GA.1b}
A_+
=
\frac{[\hat{e}_{3+},\hat{e}_{2-}]_{\mu}(b)}{\hat{W}_2^E},\
B_+
=
\frac{[\hat{e}_{2+},\hat{e}_{3+}]_{\mu}(b)}{\hat{W}_2^E}
,\
A_-
=
\frac{[\hat{e}_{1-},\hat{e}_{2-}]_{\mu}(a)}{\hat{W}_2^E},\
B_-
=
\frac{[\hat{e}_{2+},\hat{e}_{1-}]_{\mu}(a)}{\hat{W}_2^E},
\end{eqnarray}
\begin{eqnarray}
\label{eqT.GA.1c}
C_+
=
\frac{A_+[\hat{e}_{2+},\hat{e}_{1-}]_{\mu}(a)+B_+[\hat{e}_{2-},\hat{e}_{1-}]_{\mu}(a)}{\hat{W}_1^E},\
D_+
=
\frac{A_+[\hat{e}_{1+},\hat{e}_{2+}]_{\mu}(a)+B_+[\hat{e}_{1+},\hat{e}_{2-}]_{\mu}(a)}{\hat{W}_1^E},
\end{eqnarray}
\begin{eqnarray}
\label{eqT.GA.1d}
C_-
=
\frac{A_-[\hat{e}_{2+},\hat{e}_{3-}]_{\mu}(b)+B_-[\hat{e}_{2-},\hat{e}_{3-}]_{\mu}(b)}{\hat{W}_3^E},\
D_-
=
\frac{A_-[\hat{e}_{3+},\hat{e}_{2+}]_{\mu}(b)+B_-[\hat{e}_{3+},\hat{e}_{2-}]_{\mu}(b)}{\hat{W}_3^E}.
\end{eqnarray}
\end{subequations}
The boundary conditions are typically $\hat{e}_{3+}\rightarrow0,\hat{e}_{1-}\rightarrow0$ as $z\rightarrow\infty,z\rightarrow-\infty$ respectively. Here $[e_i,e_j]_{\mu}=e'_ie_j/\mu_i-e_ie'_j/\mu_j$ and the generalized Wronskians are $\hat{W}^E_i=[\hat{e}_{i+},\hat{e}_{i-}]_{\mu}$. The corresponding TM terms are obtained by making the substitutions $e\rightarrow h,\ E\rightarrow H$ and $\mu\rightarrow\varepsilon$.
\par The TE contribution to the reduced energy per unit area, with the boundary condition $e_+(\infty)=e_-(-\infty)=0$, is
\begin{eqnarray}
\label{eqT.GA.EM.1}
\int^{\infty}_{-\infty}dz\ u^E
&=&
\frac{-1}{2W^E_{\zeta,\mathbf{k}}}\frac{e'_+e_-}{\mu}\bigg|^{\infty}_{-\infty}
+
\frac{-\zeta}{2W^E_{\zeta,\mathbf{k}}}\bigg[\frac{e'_+(\zeta,\mathbf{k};z)}{\mu(\zeta,z)}\frac{\partial}{\partial\zeta}e_-(\zeta,\mathbf{k};z)
-e_+(\zeta,\mathbf{k};z)\frac{\partial}{\partial\zeta}\frac{e'_-(\zeta,\mathbf{k};z)}{\mu(\zeta,z)}\bigg]^{\infty}_{-\infty}
,
\end{eqnarray}
where the identity
\begin{eqnarray}
\label{eqT.GA.EM.2}
\frac{\partial}{\partial z}\bigg[\frac{e'_+(\zeta,\mathbf{k};z)}{\mu(\zeta,z)}\frac{\partial}{\partial\zeta}e_-(\zeta,\mathbf{k};z)
-e_+(\zeta,\mathbf{k};z)\frac{\partial}{\partial\zeta}\frac{e'_-(\zeta,\mathbf{k};z)}{\mu(\zeta,z)}\bigg]
=
-\frac{1}{\mu}\frac{\partial(\varepsilon\mu\zeta^2)}{\partial\zeta}e_+e_-+\frac{\partial\ln\mu}{\partial\zeta}\frac{\partial}{\partial z}\bigg(\frac{e'_+e_-}{\mu}\bigg)
\end{eqnarray}
has been used. The $zz$-component of the reduced stress tensor at any $z$ is
\begin{eqnarray}
\label{eqT.GA.EM.3}
t^E_{zz}(z)
=
\frac{-1}{2\mu W^E_{\zeta,\mathbf{k}}}\bigg[e'_+e'_--\varepsilon\mu\zeta^2e_+e_--k^2e_+e_-\bigg](z)
=
\frac{\partial}{\partial z_+}
\frac{[e_+,e_-]_{\mu}(z)}{2W^E}
=
-\frac{\partial}{\partial z_-}
\frac{[e_+,e_-]_{\mu}(z)}{2W^E}
,
\end{eqnarray}
where the derivatives with respect to $z_\pm$ act on the $e_\pm$ related terms, respectively. If we consider only the part depending on the position of the interfaces $z=a$ and $z=b$, we have
\begin{eqnarray}
\label{eqT.GA.EM.4}
\int^{\infty}_{-\infty}dz\ \Delta u^E
&=&
\frac{-\zeta}{2}\frac{\partial}{\partial\zeta}
\ln\Delta^E(a,b),\
\Delta^E(a,b)=[\hat{e}_{1-},\hat{e}_{2+}]_{\mu}(a)[\hat{e}_{2-},\hat{e}_{3+}]_{\mu}(b)-[\hat{e}_{1-},\hat{e}_{2-}]_{\mu}(a)[\hat{e}_{2+},\hat{e}_{3+}]_{\mu}(b)
,\quad\quad
\end{eqnarray}
and the $zz$-components of the reduced stress tensor at $z=b_-$ and $z=b_+$ satisfy
\begin{eqnarray}
\label{eqT.GA.EM.5}
t^E_{zz}(b_+)
=
\frac{1}{2}\frac{\partial\ln\Delta^E(a,b)}{\partial b_3}
,\
t^E_{zz}(b_-)
=
-\frac{1}{2}\frac{\partial\ln\Delta^E(a,b)}{\partial b_2}
\Rightarrow
t^E_{zz}(b_-)-t^E_{zz}(b_+)
=
-\frac{1}{2}\frac{\partial\ln\Delta^E(a,b)}{\partial b}
.
\end{eqnarray}
The integral over frequency and wavenumbers of this result demonstrates that the principle of virtual work is satisfied. Corresponding contributions from the TM mode are obtained by the substitutions $\varepsilon\leftrightarrow\mu$ and $\hat{e}_{i\pm}\rightarrow\hat{h}_{i\pm}$.


\section{Renormalization Scheme}
\label{appT.RS}
\par It is well known that divergences (bulk, surface, etc.)\ plague all kinds of Casimir problems. A finite Casimir force could hardly be obtained without proper subtraction of some unphysical divergences from the stress tensor and energy density of the electromagnetic field, subtraction of which are sometimes referred to as ``Lifshitz regularization" for the homogeneous cases.

\par For the Casimir force in inhomogeneous media, we propose a renormalization scheme. To extract the interaction parts, we analytically extend the material in region II to region I as shown in Fig.~\ref{Fig1}b, as the reference configuration, which would render the pressure finite.

\par For the $z=b$ interface, its reference structure consists of media in $z<b$ and $z>b$, whose permittivities and permeabilities are respectively $(\varepsilon_2,\mu_2)$ and $(\varepsilon_3,\mu_3)$. By setting $(\varepsilon_1,\mu_1)\rightarrow(\varepsilon_2,\mu_2)$, we obtain the stress tensor for this reference structure as
\begin{eqnarray}
\label{eqT.RS.4}
\tilde{t}^E_{zz}(b_-)
=
-\frac{1}{2}\frac{\partial\ln\tilde{\Delta}^E(b)}{\partial b_2},\
\tilde{t}^E_{zz}(b_+)
=
\frac{1}{2}\frac{\partial\ln\tilde{\Delta}^E(b)}{\partial b_3}
,
\end{eqnarray}
where $\tilde{\Delta}^E(b)=[\hat{e}_{3+},\hat{e}_{2-}]_{\mu}(b)$ and the boundary conditions are typically $\hat{e}_{3+}\rightarrow0,\hat{e}_{2-}\rightarrow0$ as $z\rightarrow\infty,z\rightarrow-\infty$ respectively. We propose that the renormalized stress tensors and energy densities be $\mathbf{t}_r=\mathbf{t}-\tilde{\mathbf{t}}$ and $u_r=u-\tilde{u}$, so that
\begin{eqnarray}
\label{eqT.RS.5}
t^E_{r;zz}(b_-)
=
-\frac{1}{2}\frac{\partial\ln\Delta^{E}_r(a,b)}{\partial b_2},\
t^E_{r;zz}(b_+)
=
\frac{1}{2}\frac{\partial\ln\Delta^{E}_r(a,b)}{\partial b_3},\ \Delta^{E}_r(a,b)=1-\frac{[\hat{e}_{1-},\hat{e}_{2-}]_{\mu}(a)[\hat{e}_{2+},\hat{e}_{3+}]_{\mu}(b)}{[\hat{e}_{1-},\hat{e}_{2+}]_{\mu}(a)[\hat{e}_{2-},\hat{e}_{3+}]_{\mu}(b)}.
\end{eqnarray}
Corresponding terms for the TM mode are obtained by making the substitutions $\varepsilon\leftrightarrow\mu,\ E\rightarrow H$ and $\hat{e}\rightarrow\hat{h}$.

\subsection{Homogeneous Cases}
\label{appT.RS.HC}
\par As a test of our renormalization scheme, we consider the case where each of the three slabs is homogeneous, which means the equations and solutions for each region are
\begin{eqnarray}
\label{eqT.RS.HC.1}
(\partial_z^2-\varepsilon_i\mu_i\zeta^2-k^2)(\hat{e}_{i\pm},\hat{h}_{i\pm})(\zeta,\mathbf{k};z)=0;\quad
(\hat{e}_{i\pm},\hat{h}_{i\pm})(\zeta,\mathbf{k};z)=e^{\mp\kappa_iz},
\end{eqnarray}
where $\kappa_i=\sqrt{\varepsilon_i\mu_i\zeta^2+k^2}$. Then $\Delta^E_r(a,b)$ is written as
\begin{eqnarray}
\label{eqT.RS.HC.2}
\Delta^E_r(a,b)
=
1
-
\frac{(\mu_1\kappa_2-\mu_2\kappa_1)(\mu_3\kappa_2-\mu_2\kappa_3)}{
(\mu_1\kappa_2+\mu_2\kappa_1)(\mu_3\kappa_2+\mu_2\kappa_3)}e^{-2\kappa_2(b-a)}
,\
\Delta^H_r(a,b)
=
1
-
\frac{(\varepsilon_1\kappa_2-\varepsilon_2\kappa_1)(\varepsilon_3\kappa_2-\varepsilon_2\kappa_3)}{
(\varepsilon_1\kappa_2+\varepsilon_2\kappa_1)(\varepsilon_3\kappa_2+\varepsilon_2\kappa_3)}e^{-2\kappa_2(b-a)}
.\quad\quad
\end{eqnarray}
Therefore, the TE contribution to the force per unit area is
\begin{eqnarray}
\label{eqT.RS.HC.3}
\mathcal{F}^{\TE}
&=&
-\int\frac{d\zeta d^2k}{(2\pi)^3}\frac{\kappa_2}{
d^E},\quad
d^E
=
\frac{(\mu_1\kappa_2+\mu_2\kappa_1)(\mu_3\kappa_2+\mu_2\kappa_3)}{(\mu_1\kappa_2-\mu_2\kappa_1)(\mu_3\kappa_2-\mu_2\kappa_3)}e^{2\kappa_2(b-a)}-1
,
\end{eqnarray}
and its counterpart for TM is obtained with the substitution $\mu\rightarrow\varepsilon,E\rightarrow H$. This is the DLP formula~\cite{Dzyaloshinskii1961}.

\subsection{Generalized Casimir Configuration}
\label{appT.RS.GCC}
\par Casimir's original configuration to demonstrate the measurable effect of the zero-point energy is two parallel perfectly conducting plates separated by vacuum. We generalize the Casimir configuration to the inhomogeneous case, i.e., two parallel perfectly conducting plates separated by an inhomogeneous material, by assuming that the media on the left and right sides are homogeneous and satisfy the relations $\mu_1=1,\varepsilon_1\rightarrow\infty$ and $\mu_3=1,\varepsilon_3\rightarrow\infty$. Then we have $\kappa_1,\kappa_3\sim\sqrt{\varepsilon_1},\sqrt{\varepsilon_3}\rightarrow\infty$ and the renormalized quantities $\Delta^E_r,\Delta^H_r$ are 
\begin{subequations}
\label{eqT.RS.GCC.1}
\begin{eqnarray}
\label{eqT.RS.GCC.1a}
\Delta^{E}_r(a,b)
\rightarrow
1-\frac{[\kappa_1e^{\kappa_1a}\hat{e}_{2-}(a)][\hat{e}_{2+}(b)\kappa_3e^{-\kappa_3b}]}{[\kappa_1e^{\kappa_1a}\hat{e}_{2+}(a)][\hat{e}_{2-}(b)\kappa_3e^{-\kappa_3b}]}
=
1-\frac{\hat{e}_{2-}(a)\hat{e}_{2+}(b)}{\hat{e}_{2+}(a)\hat{e}_{2-}(b)},
\end{eqnarray}
\begin{eqnarray}
\label{eqT.RS.GCC.1b}
\Delta^{H}_r(a,b)
\rightarrow
1-\frac{[-e^{\kappa_1a}\hat{h}'_{2-}(a)/\varepsilon_2(a)][e^{-\kappa_3b}\hat{h}'_{2+}(b)/\varepsilon_2(b)]}{[-e^{\kappa_1a}\hat{h}'_{2+}(a)/\varepsilon_2(a)][e^{-\kappa_3b}\hat{h}'_{2-}(b)/\varepsilon_2(b)]}
=
1-\frac{\hat{h}'_{2-}(a)\hat{h}'_{2+}(b)}{\hat{h}'_{2+}(a)\hat{h}'_{2-}(b)}
.
\end{eqnarray}
\end{subequations}
When the material in $a<z<b$ region is homogeneous, then the TE and TM contributions to the pressure on the $z=b$ interface is
\begin{eqnarray}
\label{eqT.RS.GCC.2}
\mathcal{F}^{HE}
=
\mathcal{F}^{HM}
=
-\frac{1}{2}\frac{\partial}{\partial b}
\int\frac{d\zeta d^2k}{(2\pi)^3}\ln\bigg[1-e^{-2\kappa_2(b-a)}\bigg]
=
-\frac{\pi^2}{480\sqrt{\varepsilon_2\mu_2}}\frac{1}{(b-a)^4}
.
\end{eqnarray}

\section{WKB Analysis}
\label{appT.RS.GB}
\par The point of our renormalization scheme is to achieve a finite stress and energy. To demonstrate that this is so, we employ the WKB method. For the TE mode, the WKB approximation and corresponding differential equation are
\begin{subequations}
\label{eqT.RS.GB.1}
\begin{eqnarray}
\hat{e}_i(\zeta,\mathbf{k};z)\sim\exp\bigg[\frac{1}{\epsilon}\sum_{n=0}^{\infty}\epsilon^n\hat{S}_{i,n}(\zeta,\mathbf{k};z)\bigg],\quad
\hat{e}'_i(\zeta,\mathbf{k};z)\sim\frac{1}{\epsilon}\sum_{n=0}^{\infty}\epsilon^n\hat{S}'_{i,n}(\zeta,\mathbf{k};z)\exp\bigg[\frac{1}{\epsilon}\sum_{n=0}^{\infty}\epsilon^n\hat{S}_{i,n}(\zeta,\mathbf{k};z)\bigg],
\label{eqT.RS.GB.1a}
\end{eqnarray}
\begin{eqnarray}
\hat{e}''_i(\zeta,\mathbf{k};z)\sim\bigg\{\frac{1}{\epsilon}\sum_{n=0}^{\infty}\epsilon^n\hat{S}''_{i,n}(\zeta,\mathbf{k};z)+\frac{1}{\epsilon^2}\bigg[\sum_{n=0}^{\infty}\epsilon^n\hat{S}'_{i,n}(\zeta,\mathbf{k};z)
\bigg]^2\bigg\}\exp\bigg[\frac{1}{\epsilon}\sum_{n=0}^{\infty}\epsilon^n\hat{S}_{i,n}(\zeta,\mathbf{k};z)\bigg],
\label{eqT.RS.GB.1b}
\end{eqnarray}
\begin{eqnarray}
\label{eqT.RS.GB.1c}
\bigg[\epsilon^2\partial_z^2-\epsilon\frac{\mu'_i}{\mu_i}\partial_z-\varepsilon_i\mu_i\zeta^2-k^2\bigg]\hat{e}_{i}(\zeta,\mathbf{k};z)=0.
\end{eqnarray}
\end{subequations}
The leading WKB term is
\begin{eqnarray}
\label{eqT.RS.GB.2}
\hat{S}_{i,0;\mp}(\zeta,\mathbf{k};z)
=
\frac{1}{2}\int^z_0dx\bigg(\frac{\mu'_i}{\mu_i}\pm\sqrt{\frac{\mu^{\prime2}_i}{\mu^2_i}+4\varepsilon_i\mu_i\zeta^2+4k^2}\bigg)
\sim
\pm\int^z_0dx\sqrt{\varepsilon_i\mu_i\zeta^2+k^2},
\end{eqnarray}
because the WKB solution applies for large $\zeta^2$ and $k^2$.
So the leading behavior of $\Delta^{E}_r$ for the GLC from Eq.~\eqref{eqT.RS.5} is
\begin{eqnarray}
\label{eqT.RS.GB.3}
\Delta^{E,(0)}_r(a,b)-1
&\sim&
-\frac{[\hat{S}'_{1,0;-}(a)/\mu_1(a)-\hat{S}'_{2,0;-}(a)/\mu_2(a)][\hat{S}'_{2,0;+}(b)/\mu_2(b)-\hat{S}'_{3,0;+}(b)/\mu_3(b)]}{
[\hat{S}'_{1,0;-}(a)/\mu_1(a)-\hat{S}'_{2,0;+}(a)/\mu_2(a)][\hat{S}'_{2,0;-}(b)/\mu_2(b)-\hat{S}'_{3,0;+}(b)/\mu_3(b)]}
\nonumber\\
& &
\times
\exp\bigg(-2\int^b_adx\sqrt{\varepsilon_2\mu_2\zeta^2+k^2}\bigg),
\end{eqnarray}
while the leading behavior of $\Delta^{E}_r$ for the GCC from Eq.~\eqref{eqT.RS.GCC.1a} is
\begin{eqnarray}
\label{eqT.RS.GB.4}
\Delta^{E,(0)}_r(a,b)
\sim
1-
\exp\bigg(-2\int^b_adx\sqrt{\varepsilon_2\mu_2\zeta^2+k^2}\bigg).
\end{eqnarray}
Following similar arguments, one could get the general behaviors for the TM mode contributions. It follows that the energy and hence the stress are finite according to Eq.~\eqref{eqRA.6}.

\section{Analytically Solvable Examples}
\label{appT.RS.ASE}
\par As first examples for the application of our method, we give two analytically solvable models, which illustrate our proposals.

\subsection{Inverse Square Material}
\label{appT.RS.ASE.ISM}
\par Consider the configuration where the media on the lower and upper sides, $z\leq a$ and $z\geq b$, are homogeneous and are separated by a medium in $a<z<b$ whose permittivity and permeability are $\varepsilon_2=\lambda/(c-z)^2$ and $\mu_2=1$, respectively, with $\lambda$ and $c>b$ constants. Then on the two sides $(\hat{e}_{i\pm},\hat{h}_{i\pm})(\zeta,\mathbf{k};z)=e^{\mp\kappa_iz},\ i=1,3$ and the equations to solve for the case where $\varepsilon_2$ and $\mu_2$ are extended analytically to the whole space are
\begin{eqnarray}
\label{eqT.RS.ASE.ISM.1}
\bigg[y^2\partial_y^2-\lambda\zeta^2-k^2y^2\bigg]\hat{e}_{2\pm}(\zeta,\mathbf{k};y)=0,\quad
\bigg[y^2\partial_y^2+2y\partial_y-\lambda\zeta^2-k^2y^2\bigg]\hat{h}_{2\pm}(\zeta,\mathbf{k};y)=0,
\end{eqnarray}
where $y=c-z$. The solutions are ($\nu^2=\lambda\zeta^2+1/4$)
\begin{subequations}
\label{eqT.RS.ASE.ISM.2}
\begin{eqnarray}
\label{eqT.RS.ASE.ISM.2a}
\hat{e}_{2+}(\zeta,\mathbf{k};z)=\sqrt{c-z}I_{\nu}[k(c-z)],\
\hat{e}_{2-}(\zeta,\mathbf{k};z)=\sqrt{c-z}K_{\nu}[k(c-z)],
\end{eqnarray}
\begin{eqnarray}
\label{eqT.RS.ASE.ISM.2b}
\hat{h}_{2+}(\zeta,\mathbf{k};z)=\frac{I_{\nu}[k(c-z)]}{\sqrt{c-z}},\
\hat{h}_{2-}(\zeta,\mathbf{k};z)=\frac{K_{\nu}[k(c-z)]}{\sqrt{c-z}},
\end{eqnarray}
because the $+$ solutions must be well-behaved at $z=c$, while the $-$ solutions must vanish at $-\infty$.
\end{subequations}
Therefore, $\Delta^{E}_r$ satisfies
\begin{eqnarray}
\label{eqT.RS.ASE.ISM.3}
\Delta^{E}_r(a,b)-1
&=&
-\frac{\mu_1k(c-a)K_{\nu+1}[k(c-a)]-[(\nu+1/2)\mu_1+\kappa_1\mu_2(a)(c-a)]K_{\nu}[k(c-a)]}{
\mu_1k(c-a)I_{\nu+1}[k(c-a)]+[(\nu+1/2)\mu_1+\kappa_1\mu_2(a)(c-a)]I_{\nu}[k(c-a)]
}
\nonumber\\
& &
\times
\frac{\mu_3k(c-b)I_{\nu+1}[k(c-b)]+[(\nu+1/2)\mu_3-\kappa_3\mu_2(b)(c-b)]I_{\nu}[k(c-b)]}{
\mu_3k(c-b)K_{\nu+1}[k(c-b)]-[(\nu+1/2)\mu_3-\kappa_3\mu_2(b)(c-b)]K_{\nu}[k(c-b)]
}
,
\end{eqnarray}
which means the forces per unit area at $z=b$ are
\begin{subequations}
\label{eqT.RS.ASE.ISM.4}
\begin{eqnarray}
\label{eqT.RS.ASE.ISM.4a}
\mathcal{F}^E
&=&
\frac{1}{4\pi^2(c-a)^4}\frac{\partial}{\partial\delta}\int^\infty_0d\kappa\int^{\frac{\pi}{2}}_{0}d\theta \kappa^2\sin\theta\ln\bigg\{1-
\frac{\hat{\mathcal{K}}_{\eta}(k)-\frac{\mu_2(a)}{\mu_1}\kappa_1K_{\eta}(k)}{
\hat{\mathcal{I}}_{\eta}(k)+\frac{\mu_2(a)}{\mu_1}\kappa_1I_{\eta}(k)}
\frac{\hat{\mathcal{I}}_{\eta}(k\delta)-\frac{\mu_2(b)}{\mu_3}\kappa_3\delta I_{\eta}(k\delta)}{
\hat{\mathcal{K}}_{\eta}(k\delta)+\frac{\mu_2(b)}{\mu_3}\kappa_3\delta K_{\eta}(k\delta)}\bigg\}
,
\end{eqnarray}
\begin{eqnarray}
\label{eqT.RS.ASE.ISM.4b}
\mathcal{F}^H
&=&
\frac{1}{4\pi^2(c-a)^4}\frac{\partial}{\partial\delta}\int^\infty_0d\kappa\int^{\frac{\pi}{2}}_0d\theta\kappa^2\sin\theta\ln\bigg\{1-\frac{\mathcal{K}_{\eta}(k)-\frac{\varepsilon_2(a)}{\varepsilon_1}\kappa_1K_{\eta}(k)}{
\mathcal{I}_{\eta}(k)+\frac{\varepsilon_2(a)}{\varepsilon_1}\kappa_1I_{\eta}(k)}
\frac{\mathcal{I}_{\eta}(k\delta)-\frac{\varepsilon_2(a)}{\varepsilon_3}\frac{\kappa_3}{\delta}I_{\eta}(k\delta)}{
\mathcal{K}_{\eta}(k\delta)+\frac{\varepsilon_2(a)}{\varepsilon_3}\frac{\kappa_3}{\delta}K_{\eta}(k\delta)}\bigg\}
,
\end{eqnarray}
\end{subequations}
where $k$ and $\kappa$ are rescaled to dimensionless form,  $(\hat{\mathcal{I}}_{\eta},\mathcal{I}_{\eta})(x)=xI_{\eta+1}(x)+(\eta\pm1/2)I_{\eta}(x),(\hat{\mathcal{K}}_{\eta},\mathcal{K}_{\eta})(x)=xK_{\eta+1}(x)-(\eta\pm1/2)K_{\eta}(x)$, $\delta=(c-b)/(c-a)\in(0,1)$, $k=\kappa\sin\theta,\zeta=\kappa\cos\theta$, and $\eta=\sqrt{\varepsilon_2(a)\zeta^2+1/4}$. For the generalized Casimir configuration limit, i.e. $\mu_1=\mu_3=1,\ \varepsilon_1,\varepsilon_3\rightarrow\infty$, we have
\begin{subequations}
\label{eqT.RS.ASE.ISM.5}
\begin{eqnarray}
\label{eqT.RS.ASE.ISM.5a}
\mathcal{F}^E
&=&
-\frac{1}{4\pi^2(c-a)^4\delta}\int^\infty_0d\kappa\int^{\frac{\pi}{2}}_{0}d\theta \kappa^2\sin\theta
\frac{K_{\eta}(k)/K_{\eta}(k\delta)}{I_{\eta}(k)K_{\eta}(k\delta)-K_{\eta}(k)I_{\eta}(k\delta)}
,
\end{eqnarray}
\begin{eqnarray}
\label{eqT.RS.ASE.ISM.5b}
\mathcal{F}^H
&=&
-\frac{1}{4\pi^2(c-a)^4\delta}\int^\infty_0d\kappa\int^{\frac{\pi}{2}}_{0}d\theta \kappa^2\sin\theta
\frac{(k^2\delta^2+\eta^2-1/4)\mathcal{K}_{\eta}(k)/\mathcal{K}_{\eta}(k\delta)
}{\mathcal{I}_{\eta}(k)\mathcal{K}_{\eta}(k\delta)-\mathcal{K}_{\eta}(k)\mathcal{I}_{\eta}(k\delta)}
.
\end{eqnarray}
\end{subequations}


\subsection{Diaphanous Material}
\label{appT.RS.ASE.DM}
\par Rewriting $e,h$ as $e_{\pm}(\zeta,\mathbf{k};z)=\sqrt{\mu}p_{\pm}(\zeta,\mathbf{k};z)$ and $h_{\pm}(\zeta,\mathbf{k};z)=\sqrt{\varepsilon}q_{\pm}(\zeta,\mathbf{k};z)$, we find the equations of motion to be
\begin{eqnarray}
\bigg[\partial_z^2-\varepsilon\mu\zeta^2-k^2+\frac{\mu''}{2\mu}-\frac{3\mu^{\prime2}}{4\mu^2}\bigg]p_{\pm}(\zeta,\mathbf{k};z)=0,\
\bigg[\partial_z^2-\varepsilon\mu\zeta^2-k^2+\frac{\varepsilon''}{2\varepsilon}-\frac{3\varepsilon^{\prime2}}{4\varepsilon^2}\bigg]q_{\pm}(\zeta,\mathbf{k};z)=0.
\label{eqT.RS.ASE.DM.1}
\end{eqnarray}

\par Consider the GCC for a diaphanous material, which satisfies $\varepsilon\mu=1$. For the particular case $\varepsilon=e^{\lambda(z-c)^2}$, where $\lambda$ is a nonzero constant, the equations for the TE and TM modes are
\begin{eqnarray}
p''_{\pm}(y)
+
\bigg[-\frac{\kappa^2}{2\lambda}-1+\frac{1}{2}-\frac{y^2}{4}\bigg]p_{\pm}(y)=0,\quad
q''_{\pm}(y)
+
\bigg[-\frac{\kappa^2}{2\lambda}+\frac{1}{2}-\frac{y^2}{4}\bigg]q_{\pm}(y)=0,
\label{eqT.RS.ASE.DM.2}
\end{eqnarray}
where $y=\sqrt{2\lambda}(z-c)$ and $p(z)=p(y),\ q(z)=q(y)$. So $\hat{e}$ and $\hat{h}$ are exactly solved as
\begin{eqnarray}
\hat{e}_{\pm}(z)
=
e^{-\frac{y^2}{4}}\DD_{-\frac{\kappa^2}{2\lambda}-1}(\pm y),\quad
\hat{h}'_{\pm}(z)
=
\sqrt{2\lambda}\frac{d}{dy}e^{\frac{y^2}{4}}\DD_{-\frac{\kappa^2}{2\lambda}}(\pm y)
=
\sqrt{2\lambda}e^{\frac{y^2}{4}}[y\DD_{-\frac{\kappa^2}{2\lambda}}(\pm y)\mp\DD_{1-\frac{\kappa^2}{2\lambda}}(\pm y)].
\label{eqT.RS.ASE.DM.3}
\end{eqnarray}
where $\DD_{\nu}(x)$ is the parabolic cylinder function. The pressures on the $z=b$ interface are the same for the TE and TM modes:
\begin{eqnarray}
\label{eqT.RS.ASE.DM.4}
\mathcal{F}^E
=
\mathcal{F}^H
&=&
-\frac{\lambda^{2}}{\pi^2}\int^{\infty}_0d\kappa\kappa^2\frac{\frac{\DD_{-\kappa^2-1}(-y_a)}{\DD_{-\kappa^2-1}(-y_b)}
[\DD_{-\kappa^2-1}(y_b)\DD_{-\kappa^2}(-y_b)+\DD_{-\kappa^2-1}(-y_b)\DD_{-\kappa^2}(y_b)]
}{\DD_{-\kappa^2-1}(y_a)\DD_{-\kappa^2-1}(-y_b)-\DD_{-\kappa^2-1}(-y_a)\DD_{-\kappa^2-1}(y_b)}
,
\end{eqnarray}
where $\kappa$ is rescaled to dimensionless form and $y_z=\sqrt{2\lambda}(z-c)$.

\acknowledgments
We thank the U.S. National Science Foundation, Grant
No. 1707511, for support of this research. We acknowledge Prachi Parashar, Alex Mau, Hannah Day, and Pushpa Kalauni for helpful discussions. GK thanks the Department of Mathematics at Texas A\&M University for its kind hospitality during part of this research.

\bibliography{ref}

\end{document}